\documentclass[english]{aa}
\usepackage{mathptmx}
\usepackage[T1]{fontenc}
\usepackage[latin9]{inputenc}
\usepackage{array}
\usepackage{verbatim}
\usepackage{textcomp}
\usepackage{amsmath}
\usepackage{graphicx}

\makeatletter

\DeclareRobustCommand*{\lyxarrow}{%
\@ifstar
{\leavevmode\,$\triangleleft$\,\allowbreak}
{\leavevmode\,$\triangleright$\,\allowbreak}}
\newcommand{\lyxmathsym}[1]{\ifmmode\begingroup\def\b@ld{bold}
  \text{\ifx\math@version\b@ld\bfseries\fi#1}\endgroup\else#1\fi}

\providecommand{\tabularnewline}{\\}

\makeatother

\usepackage{babel}

\begin{document}

\title{Black Holes and Galactic Density Cusps}

\subtitle{From Black Hole to Bulge}

\author{M. Le Delliou\inst{1}\and R.N. Henriksen\inst{2}\and J.D. MacMillan\inst{3}}

\institute{Instituto de Física Teórica UAM/CSIC, Facultad de Ciencias, C-XI,
Universidad Autónoma de Madrid\\
 Cantoblanco, 28049 Madrid SPAIN\\
\email{Morgan.LeDelliou@uam.es}\and Queen's University, Kingston,
Ontario, Canada \\
\email{henriksn@astro.queensu.ca}\and Faculty of Science, University
of Ontario Institute of Technology, Oshawa, Ontario, Canada L1H 7K4\\
\email{joseph.macmillan@gmail.com}}

\offprints{MLeD\hfill{}Preprint: IFT-UAM/CSIC-09-28}

\date{Submitted ...; Received ...; Accepted...}

\date{}

\abstract{%------------------context (optional: leave void)
} {%-----------------------aims
In this paper we continue our study of density cusps that may contain
central black holes.% The actual coeval self-similar growth would not distinguish between the %central object and the surroundings.
 } {%-----------------------methods
%To study the environment of an existing black hole we seek descriptions of %steady `cusps' that may contain a black hole and that retain at least a %memory of self-similarity. We refer to the environment in brief as the %`bulge' or sometimes the `halo'. This depends on whether the black hole is a %true singularity or rather only a core mass concentration.
We recall our attempts to use distribution functions with a memory
of self-similar relaxation, but mostly they apply only in restricted
regions of the global system. We are forced to consider related distribution
functions that are steady but not self-similar. } {%-----------------------results
%We find simple descriptions of simulated collisionless matter in the process %of examining the presence and growth of central masses. However o
One remarkably simple distribution function that has a filled loss
cone describes a bulge that transits from a near black hole domain
to an outer `zero flux' regime where $\rho\propto r^{-7/4}$. The
transition passes from an initial inverse square profile through a
region having a $1/r$ density profile. The structure is likely to
be developed at an early stage in the growth of a galaxy. A central
black hole is shown to grow exponentially in this background with
an e-folding time of a few million years. } {%-----------------------conclusions (optional: leave void) (to be structured?)
We derive our results from first principles, using only the angular
momentum integral in spherical symmetry. The initial relaxation probably
requires bar instabilities and %assuming either self-similar virialisation or normal steady virialisation. %The implied energy relaxation of the collisionless matter is due to the time %dependence. Phase mixing relaxation may be enhanced by
 clump-clump interactions.}

\keywords{theory-dark matter-galaxies:haloes-galaxies:nuclei-black hole physics-gravitation.}

\maketitle

\section{Introduction}

We discussed the relation between the formation of black holes (hereafter
BH) and of galactic bulges, or of bulges with a central mass in the
previous papers of this series of papers (Henriksen et al., referred
to here as paper \cite{HLeDMcM09a,HLeDMcM09b}, and references hereafter).

Previous research  (\cite{KR1995,Ma98}) and more recently (\cite{FM2000,Geb2000})
has established a strong correlation between the BH mass and the surrounding
stellar bulge mass (or velocity dispersion), which we take as an indication
of coeval growth. Such growth may occur as BH `seeds' accrete gas
or disrupted stars in a dissipative fashion during the AGN (Active
Galactic Nuclei) phase, but there is as yet no generally accepted
scenario. Moreover, the necessity for a seed and observations of early
very supermassive BHs (e.g. \cite{Kurk2007}), together with changes
in the normalization of the BH mass-bulge mass proportionality (e.g.
\cite{Mai2007})\textbf{;} suggest an alternate growth mechanism.
In fact some authors (\cite{PFP2008}) have studied the possible size
of the dark matter component in BH masses and deduce that between
1\% and 10\% of the black hole mass could be due to dark matter.

We explored coeval growth in the case of spherically symmetric radial
and non-radial infall (paper \cite{HLeDMcM09a,HLeDMcM09b}), using
distribution functions with a self-similar memory. However although
we obtained reasonable descriptions in sections, we did not find a
distribution function (DF) that could be fit to Black Hole halo and
bulge continuously. Moreover the problem of the black hole feeding
was left unresolved. This requires an anisotropic DF with a filled
loss cone of the nature of the Bahcall \& Wolf solution (\cite{BW76}).

Here we repair these omissions for a special case of non self-similar
infall. We use the same technique of inferring reasonable distribution
functions for collisionless matter from the time dependent Collisionless
Boltzmann (CBE) and Poisson set, while insisting on a central point
mass. The temporal evolution allows for relaxation of collisionless
matter in addition to possible `clump-clump' (two clump) interactions.

We use, in this paper as in the previous (\cite{HLeDMcM09a,HLeDMcM09b}),
the Carter-Henriksen (\cite{CH91}) procedure. In this way we obtain
a quasi-self-similar system of coordinates (\cite{H2006,H2006A})
that enables expression of the CBE-Poisson set with explicit reference
to a previous transient self-similar dynamical relaxation. Thus we
can remain `close' to self-similarity just as the simulations appear
to do.

We begin the next section with a summary of this approach in spherical
symmetry that is common to all papers in this series, including the
useful results found earlier (papers \cite{HLeDMcM09a,HLeDMcM09b}).
Subsequently we recall the limiting case with a high binding energy
cut off in the distribution function (hereafter DF). This is to remind
us that such a cut-off, existing for \emph{any} reason, produces a
flat cusp.\textbf{ }Finally we give the key result of this paper as
a non-self-similar DF (of a type found previously (\cite{EA2006})
but not used in this connection) that describes the transition zone.
This zone extends from a central mass through a shallow region (to
be compared with observations of the milky way) into a Bahcall \&
Wolf feeding region. Ultimately it enters an inverse square law region.

\section{Dynamical Equations and radial results }

Following the formulation of \cite{H2006} we transform to infall
variables the collisionless Boltzmann and Poisson equations for a
spherically symmetric anisotropic system in the `Fujiwara' form (e.g.
\cite{Fujiwara}) namely

\begin{eqnarray}
 &  & \frac{\partial f}{\partial t}+v_{r}\frac{\partial f}{\partial r}+\left(\frac{j^{2}}{r^{3}}-\frac{\partial\Phi}{\partial r}\right)\frac{\partial f}{\partial v_{r}}=0,\label{eq:Boltzmann}\\
 &  & \frac{\partial}{\partial r}\left(r^{2}\frac{\partial\Phi}{\partial r}\right)=4\pi^{2}G\int f(r,v_{r},j^{2})dv_{r}dj^{2},\label{eq:Poisson}\end{eqnarray}

where $f$ is the phase-space mass density, $\Phi$ is the `mean'
field gravitational potential, $j^{2}$ is the square of the specific
angular momentum and other notation is more or less standard.

The transformation to infall variables has the form (e.g. \cite{H2006})

\textbf{\begin{align}
R & =r\, e^{-\alpha T/a},\hspace{2cm}Y=v_{r}e^{-(1/a-1)\alpha T},\nonumber \\
Z & =j^{2}e^{-(4/a-2)\alpha T},\hspace{0.9cm}e^{\alpha T}=\alpha t,\nonumber \\
P\left(R,Y,Z;T\right) & =e^{(3/a-1)\alpha T}\pi f\left(r,v_{r},j^{2};t\right),\label{eq:artrans}\\
\Psi\left(R;T\right) & =e^{-2(1/a-1)\alpha T}\Phi(r),\nonumber \\
\Theta\left(R;T\right) & =\rho(r,t)e^{-2\alpha T}.\nonumber \end{align}
}

The passage to the self-similar limit requires taking $\partial_{T}=0$
when acting on the transformed variables. Thus the self-similar limit
is a stationary system in these variables, which is a state that we
refer to as `self-similar virialisation' (\cite{HW99}, \cite{LeD2001}).
The virial ratio $2K/|W|$ is a constant in this state (although greater
than one; $K$ is kinetic energy and $W$ is potential), but the system
is not steady in physical variables as infall continues. 

The single quantity $a$ is the constant that determines the dynamical
similarity, called the self-similar index. It is composed of two separate
reciprocal scalings, $\alpha$ in time and $\delta$ in space, in
the form $a\equiv\alpha/\delta$. As it varies it contains all dominant
physical constants of mass, length and time dimensions, since the
mass scaling $\mu$ has been reduced to $3\delta-2\alpha$ in order
to maintain Newton's constant $G$ invariant (e.g. \cite{H2006}).

We assume that time, radius, velocity and density are measured in
fiducial units $r_{o}/v_{o}$,$r_{o}$, $v_{o}$ and $\rho_{o}$ respectively.
The unit of the distribution function (DF from now on) is $f_{o}$
and that of the potential is $v_{o}^{2}$. We remove constants from
the transformed equations by taking \begin{equation}
f_{o}=\rho_{o}/v_{o}^{3},~~~~~~v_{o}^{2}=4\pi G\rho_{o}r_{o}^{2}.\label{eq:units}\end{equation}

These transformations convert equations (\ref{eq:Boltzmann}),(\ref{eq:Poisson})
to the respective forms

\begin{multline}
\frac{1}{\alpha}\partial_{T}P-(3/a-1)P+(\frac{Y}{\alpha}-\frac{R}{a})\partial_{R}P\\
-\left((1/a-1)Y+\frac{1}{\alpha}\left(\frac{\partial\Psi}{\partial R}-\frac{Z}{R^{3}}\right)\right)\partial_{Y}P-(4/a-2)Z\partial_{Z}P=0\label{eq:SSBoltzmann}\end{multline}
 and \begin{equation}
\frac{1}{R^{2}}\frac{d}{dR}\left(R^{2}\frac{\partial\Psi}{\partial R}\right)=\Theta.\label{eq:SSPoisson}\end{equation}
 This integro-differential system is closed by \begin{equation}
\Theta=\frac{1}{R^{2}}\int~PdY~dZ.\label{eq:dmoment}\end{equation}

This completes the formalism that we will use to obtain our results.

For easy reference in this paper, we summarize the general forms of
the self-similar DFs that we discussed previously.

In paper \cite{HLeDMcM09a} we discussed the emergence in purely radial
infall of the rigorously steady DFs from Henriksen and Widrow (hereafter
HWDF, \cite{HW95})\begin{equation}
\pi f=\widetilde{F}(\kappa)|E|^{1/2}\delta(j^{2}),\label{eq:steadyF}\end{equation}
and the time-dependent DF from Fridmann and Polyachenko (hereafter
FPDF, \cite{FP1984})\begin{equation}
f=\frac{K}{(-E+E_{o})^{1/2}}\delta(j^{2}).\label{eq:FPDF}\end{equation}
Introducing anisotropies in paper \cite{HLeDMcM09b}, we found the
general self-similar solution to equation (\ref{eq:SSBoltzmann})
to be \begin{equation}
P=\widetilde{P}(j^{2},\kappa)\mathcal{E}^{q},\label{eq:genanisop}\end{equation}
 where \begin{eqnarray}
\kappa & \equiv & \mathcal{E}Z^{-\frac{1-a}{2-a}},\label{eq:defs}\\
q & \equiv & \frac{3-a}{2(a-1)}.\end{eqnarray}

This yields the physical form of the general steady state with self-similar
memory to be \begin{eqnarray}
\pi f & = & \widetilde{P}(\kappa)|E|^{q},\label{eq:anisopsteady}\\
\kappa & = & |E|(j^{2})^{-(\frac{1-a}{2-a})},\nonumber \end{eqnarray}
 where $E\equiv v_{r}^{2}/2+j^{2}/(2r^{2})+\Phi$. Taking first $\widetilde{P}$
to be constant, and then taking it to be proportional to $\kappa^{-q}$.
this yields respectively \begin{eqnarray}
\pi f & = & K|E|^{q},\label{eq:DFE}\\
\pi f & = & \frac{K}{(j^{2})^{w}},\label{eq:DFJ}\end{eqnarray}
 where we have defined $w=(3-a)/(4-2a)$. These limits were found
to apply to inner and outer extremes of the relaxed region of a simulated
`bulge' (i.e. no black hole was included in the simulations). 

The DF (\ref{eq:anisopsteady}) can be put in a form that seems to
generalize the FPDF (Eq. \ref{eq:FPDF}). We choose $\widetilde{P}\propto\kappa^{-\left(\frac{1}{a-1}\right)}$
to obtain \begin{equation}
\pi f=\frac{K}{(j^{2})^{\frac{1}{2-a}}|E_{o}-E|^{1/2}}.\label{eq:genFPDF}\end{equation}
The density and potential laws are more general than those of the
FPDF, being respectively $\rho\propto r^{-2a}$ and $\Phi\propto r^{2(1-a)}$.

None of these DFs can be said to apply to every region of the simulated
halos.

\section{High Binding Energy Cut-off}

We recall in this section an extreme case that limits the flattening
of the cusp near a black hole. There is evidence (\cite{LeD2001},
\cite{MWH2006}) for a cut-off in the number of particles with high
binding energy, even without a central point mass. The presence of
a central black hole of mass $M_{\bullet}$ is likely to accentuate
this trend by accretion. We may imitate such a cut-off in order to
study limiting behaviour near the black hole, by using the isothermal
DF (see e.g. papers \cite{HLeDMcM09a} and \cite{HLeDMcM09b}) at
negative energies and temperatures in the form \begin{equation}
\pi f=Ke^{(\beta E)}.\end{equation}
 Here $E=v_{r}^{2}/2+j^{2}/(2r^{2})+\Phi<0$. We suppose that the
constant $\beta>0$ and corresponds to some reciprocal cut-off energy.
A straight-forward calculation of the density implied by such a DF
yields (note that we integrate only over negative energies rather
than over all velocities) \begin{equation}
\rho=\frac{8\sqrt{2}K}{3}|\Phi|^{3/2}M(1,5/2,-\beta|\Phi|),\end{equation}
 Where $M(a,b,z)$ is the Kummer function. We expect this to hold
where $|\Phi|\approx GM_{\bullet}/r$ is large, and in this limit
the Kummer function varies as $-3/(2z)$. Consequently the density
cusp becomes as flat as $\sqrt{|\Phi|}\propto r^{-1/2}$. This was
already noticed in (\cite{NM99}) and in (\cite{MS2006}), where in
both cases it is due to scouring by merging black holes. In any case,
as was recognized in (\cite{NM99}), only the cut-off is required.

A cut-off in the number of particles per unit energy $dM/dE$ is to
be expected on general grounds. The density of states (\cite{BT1987})
for an isotropic DF in a region dominated by a point mass potential
is $g(E)=\sqrt{2}(\pi GM_{\bullet})^{3}|E|^{-5/2}$, from which the
differential energy distribution $dM/dE=g(E)f(E)$ may be calculated.
Thus even a constant $f(E)$ (for which $\rho\propto|\Phi|^{3/2}$
and therefore goes like $r^{-1.5}$ near a point mass) will show a
cutoff in the mass distribution as the density of states decreases
in the expanding phase space. However the evidence suggests that $f(E)$
is also declining with increasing $|E|$ in the most tightly bound
central regions.

Assuming the DF (\ref{eq:DFE}) and $a=0.5$), we predict the mass
distribution function $dM/dE\propto|E|^{-5}$. Thus a cut-off in the
isotropic DF can be associated with the $r^{-2a}~=~r^{-1}$ density
profile. The mechanism for the high binding energy cut-off is not
immediately evident, but it must be part of the dynamical relaxation.
High negative energy particles must be preferentially excited to less
negative energies, which does not seem to occur in strict shell code
simulations (\cite{HW99}). However it may occur due to the presence
of sub-structure such as clumps or bar formation by the radial orbit
instability (\cite{MWH2006}). In any case it may not require black
hole `scouring', although this process does have the maximum effect
(\cite{MS2006,NM99}).

\section{Global Solution with Black hole and a Bahcall/Wolf Outer Cusp}

We consider an exact collisionless cusp that has an embedded central
mass. It is necessarily not self-similar. We have seen that we expect
a DF that is primarily dependent on angular momentum in the outer
regions of a bulge (e.g. (\ref{eq:DFJ}) before the maximum in angular
momentum). We can not follow analytically the development of this
region in time, as this is the province of numerical work. We can
however hope to find analytic descriptions of ultimate states.

There is one steady solution that is not self-similar but which is
closely related to the Evans and An (\cite{EA2006}) solution with
anisotropy parameter $\beta=0.5$. This is an exact solution to equation
(\ref{eq:SSBoltzmann}) with $\partial_{T}\ne0$ (compare equations
\ref{eq:genanisop}, \ref{eq:DFJ}) in the form (recall that $w\equiv(3-a)/(4-2a)$)
\[
P=\widetilde{P}(j^{2})Z^{-w},\]
\begin{figure*}
~\vspace{-1.4cm}

\hfill{}\quad{}\begin{tabular}{>{\centering}m{0.6\columnwidth}>{\centering}m{0.6\columnwidth}}
\includegraphics[bb=14bp 14bp 360bp 365bp,width=0.5\columnwidth]{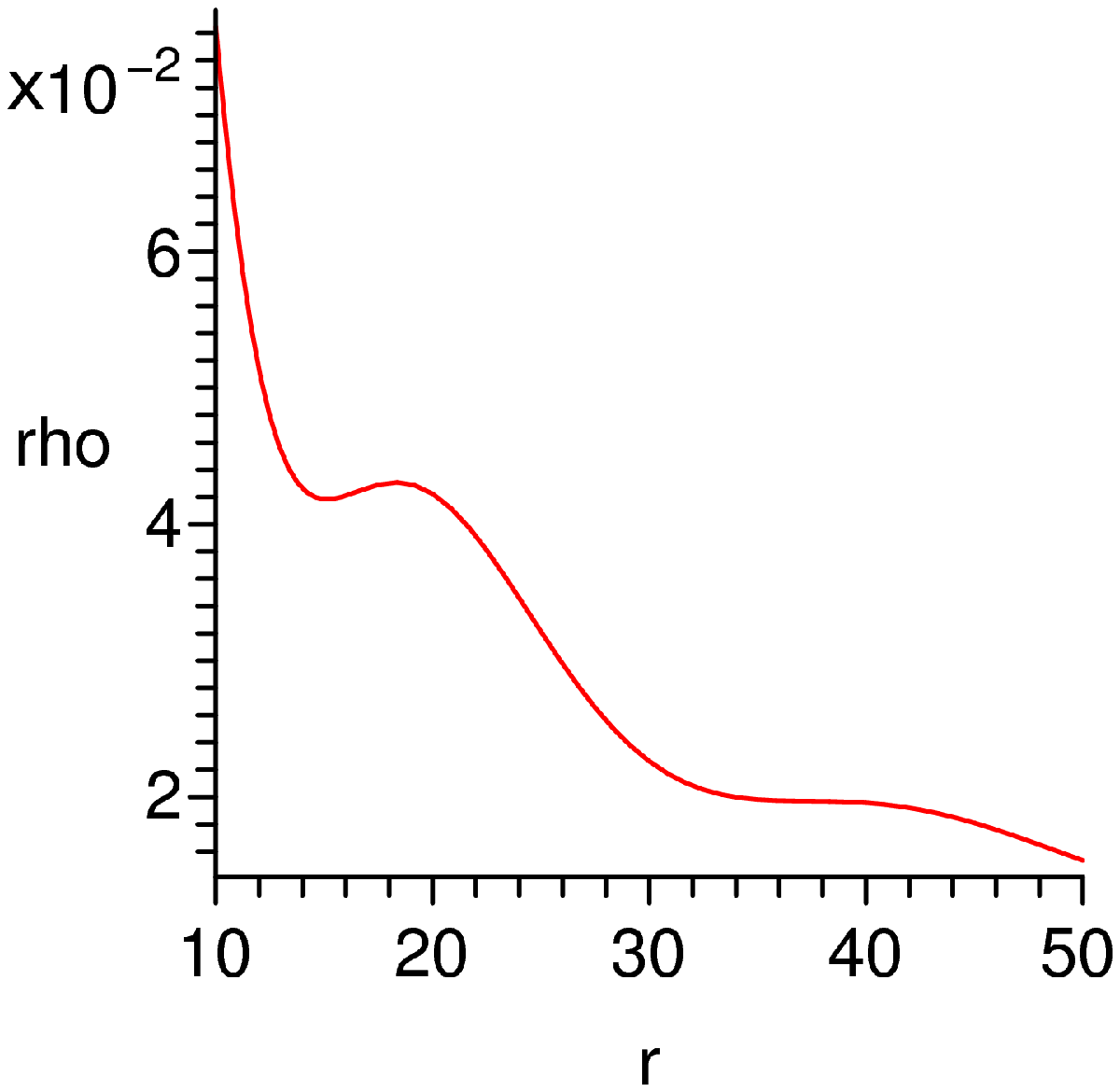} & \includegraphics[bb=20bp 118bp 575bp 673bp,width=0.75\columnwidth]{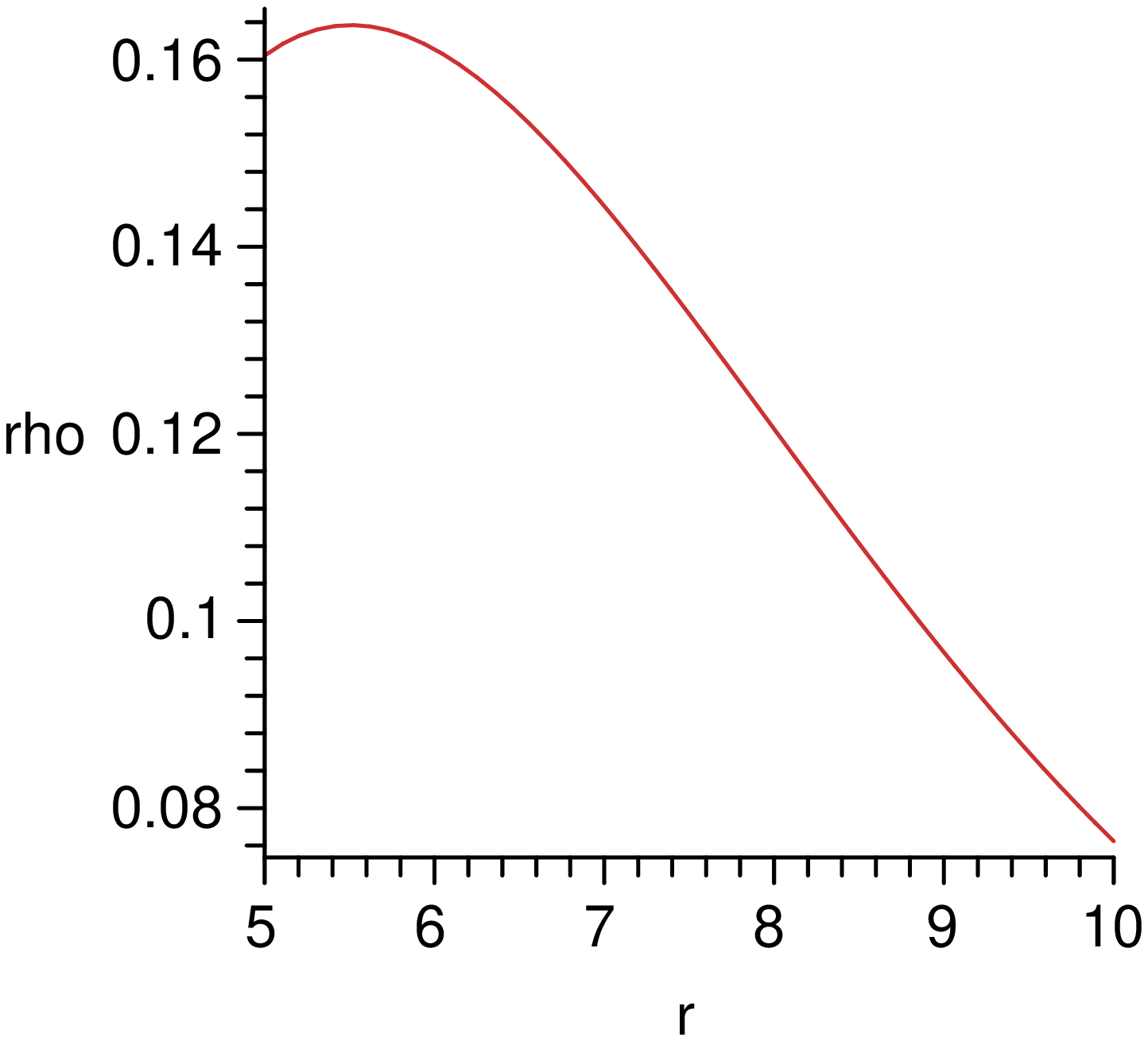}\vspace{1.6cm}
\tabularnewline[-3.001cm]
\includegraphics[bb=20bp 208bp 575bp 653bp,width=0.7\columnwidth]{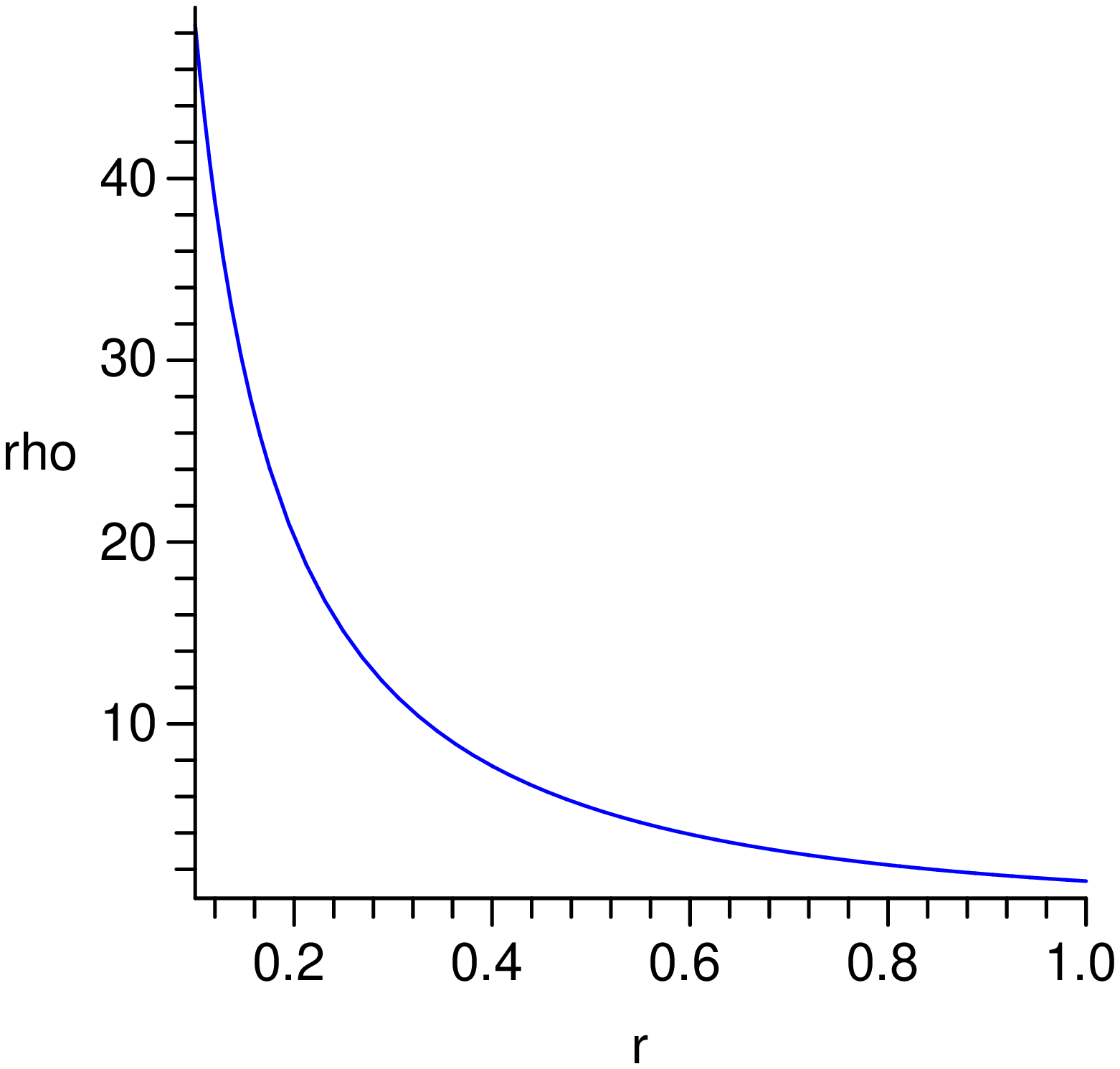} & \includegraphics[bb=20bp 218bp 575bp 673bp,width=0.7\columnwidth]{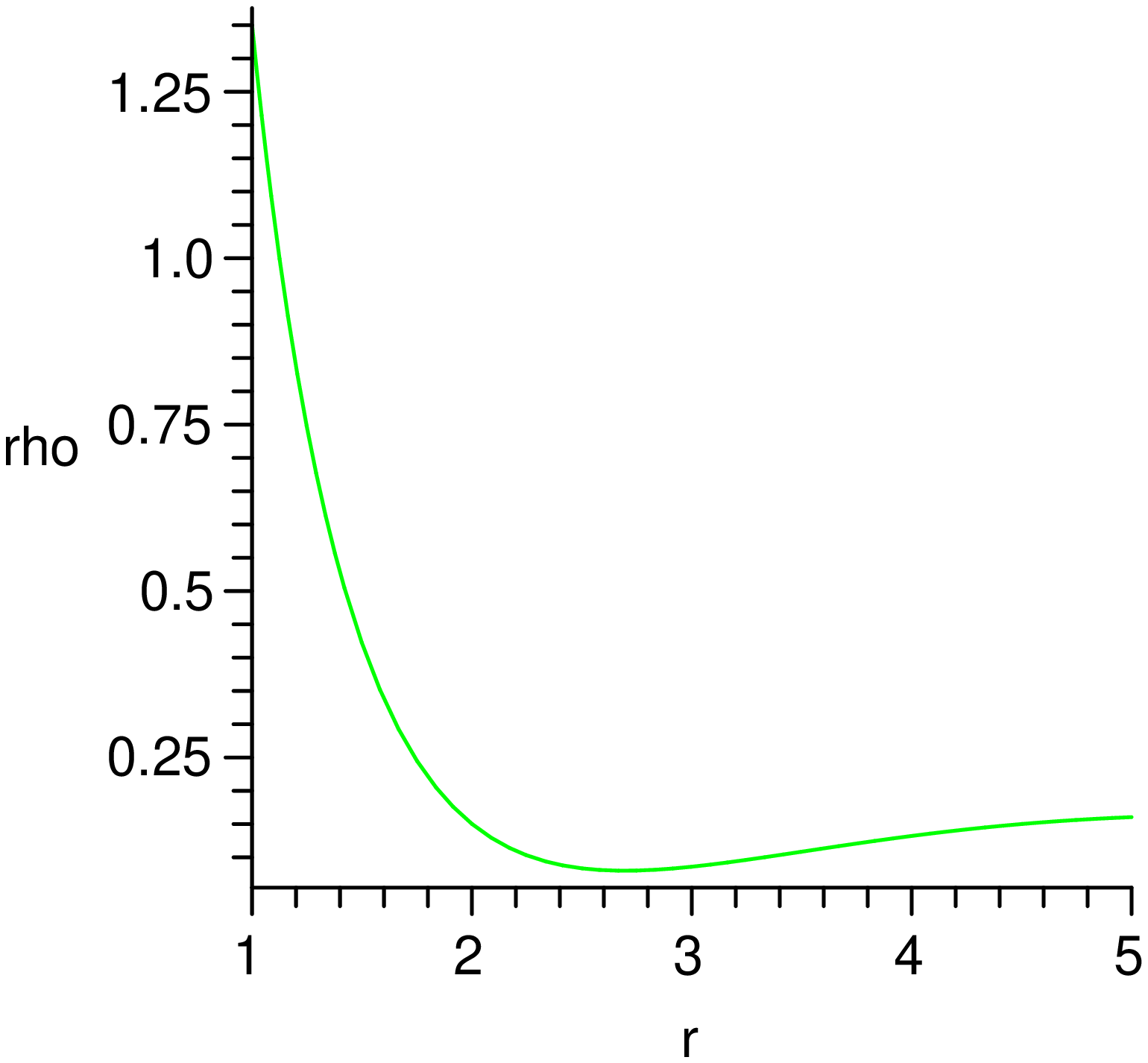}\tabularnewline
\end{tabular}\quad{}\hfill{}\vspace{1cm}

\caption{We show the various density regimes of the zeroflux cusp with an embedded
black hole. The constants in equation (\ref{eq:nonSSpot}) have been
chosen for numerical convenience and positive definiteness to be:
$|\Phi|_{\infty}=3$, $C_{1}=5$ and $C_{2}=-2/(\pi K')$. Here $K'\equiv K/(8\pi)=10$
so that $C_{2}\equiv-M_{\bullet}=-1/5\pi$. In addition the plot is
of $\rho/\sqrt{K'}$. All quantities are expressed in fiducial units;
$r_{o}$, $\rho_{o}$ and $v_{o}$. %For this example the bulge is about ten times the mass of the black hole %in the inner region. The constant is  in fiducial units.
 }

\label{fig:regimes} 
\end{figure*}
that is, written entirely in transformed coordinates, \begin{equation}
P=\widetilde{P}(Z\exp{(4/a-2)\alpha T})Z^{-w}.\end{equation}
 This solution is always steady but not always self-similar because
of the $T$ dependence. This equation is seen to satisfy equation
(\ref{eq:SSBoltzmann}) by direct substitution when the dependence
on T is included. The steady form (\ref{eq:genanisop}) that we found
in the previous paper of this series is distinguished from the general
steady class by being a possible limit of self-similar evolution.
The same problem has been addressed with arbitrary steady and isotropic
distribution functions in an interesting paper by \cite{Baes05}.

A convenient choice is to take $\widetilde{P}=K(j^{2})^{\nu}$, where
$\nu$ is any suitable real number. Then the DF is simply \begin{equation}
\pi f=K(j^{2})^{(\nu-w)}.\label{eq:nonSSDF}\end{equation}

The density associated with such a distribution is ($E<0$, $\nu-w>-1$)
\begin{equation}
\rho=\sqrt{\pi}K\frac{\Gamma(\nu-w+1)}{(\nu-w+3/2)\Gamma(\nu-w+3/2)}r^{2(\nu-w)}|2\Phi|^{(\nu-w+3/2)}.\label{eq:nonSSrho}\end{equation}
 This may now be substituted into the Poisson equation to give an
equation for $|\Phi|$ and hence the density in a self-consistent
bulge.

There is a power law solution $\propto r^{h}$ provided that $h>-1$.
In terms of $w$ and $\nu$, $h$ is \begin{equation}
h=\frac{-2(1+\nu-w)}{\nu-w+1/2},\end{equation}
 and the logarithmic density slope is $h-2$. Most of the density
power laws ($h-2$) of interest arise for values of $\nu-w$ near
$-1$. 

For example $\nu-w=-0.9$ yields $h-2=-1.5$, while $\nu-w=-5/6$
yields $h-2=-1.0$. The limiting case with $\nu-w=-1$ (which has
a logarithmic singularity at zero angular momentum) gives as expected
$h-2=-2$. Setting $\nu=0$ returns us to the self-similar bulge (\ref{eq:DFJ})
and indeed $h-2=-2a$ in this case.

It is clear however that there is a special case when $\nu-w+1/2=0$,
for which the DF is $\pi f=K/\sqrt{j^{2}}$. This value treated separately
yields \begin{equation}
\rho=\frac{2\pi K}{r}|\Phi|.\label{eq:nonSSdens}\end{equation}
 The Poisson equation is therefore linear and it is readily solved
to find \begin{equation}
|\Phi|=\sqrt{\frac{1}{2\pi K}}C_{1}\frac{J_{1}(\sqrt{8\pi Kr})}{r^{1/2}}+\pi\sqrt{2\pi K}C_{2}\frac{Y_{1}(\sqrt{8\pi Kr})}{r^{1/2}}+|\Phi|_{\infty},\label{eq:nonSSpot}\end{equation}
 where $J_{1}$ and $Y_{1}$ are first order Bessel functions of the
first and second kinds respectively. The density is given by equation
(\ref{eq:nonSSdens}). 

The constants in equation (\ref{eq:nonSSpot}) must be such as to
maintain $|\Phi|$ and $\rho$ positive despite the oscillations in
the Bessel functions%
\begin{comment}
, even though $\Phi$ is negative
\end{comment}
{}. This may impose a limited range in radius, which is in any case
required to ensure a finite mass.\textbf{ }In general the Bessel functions
will be out of phase by $\pi/2$ and it is not possibile to keep their
sum positive everywhere. However it is possible to keep the whole
expression positive by choosing an appropriate set of constants as
we have done in our example. Because the term containing the Bessel
functions is declining as $r^{-7/4}$, once a critical transition
region is traversed successfully by a choice of constants the positivity
is assured henceforward. It is really only the difference of the potential
moduli that is physical and this difference is free to oscillate as
indicated. It is interesting nevertheless that $|\Phi|_{\infty}$
is necessarily non-zero, as this suggests that the solution must be
embedded in distant matter.

As $z\rightarrow0$ one has $J_{1}(z)\approx z/2$ while $Y_{1}(z)\approx-2/(\pi z)$.
Hence the potential contains a central point mass in this limit if\textbf{
$C_{2}=-M_{\bullet}$.} The term in $J_{1}$ tends to a constant and
is presumably associated with the bulge mass itself. Consequently
by equation (\ref{eq:nonSSdens}) the density cusp near the central
black hole has the profile \begin{equation}
\rho=2\pi K\left[\frac{C_{1}+|\Phi|_{\infty}}{r}+\frac{GM_{\bullet}}{r^{2}}\right].\label{eq:CuspR1}\end{equation}
 The constant $C_{1}$ gives the difference in the modulus of the
bulge potential, coming from infinity to a radius where $J_{1}$ is
well approximated by the small argument limit (it depends on $K$).
This happens before the small argument limit applies to $Y_{1}$.
We see in general from this solution that the near black hole cusp
can start with a ($-1$) logarithmic density slope and steepen to
($-2$) near the black hole. This appears to encompass the observations.

A remarkable behaviour of this solution occurs at large $r$ (actually
large $\sqrt{8\pi Kr}$), where $J_{1}(z)\asymp\sqrt{2/(\pi z)}\cos{(z-3\pi/4)}$
and $Y_{1}(z)\asymp\sqrt{2/(\pi z)}\sin{(z-3\pi/4)}$. This shows
that the mean density tends to the $r^{-7/4}$ density profile that
is characteristic of the Bahcall and Wolf (\cite{BW76}) zero flux
solution. According to the small $r$ limits, we find it here outside
a flatter region $r^{-1}$ that gives way ultimately to an inner $r^{-2}$
region. Outside of the Bahcall/Wolf $r^{-7/4}$ region one would return
to a $r^{-1}$ profile (recalling that the potential there, i.e. at
`infinity' is not zero). This appears to connect a black hole through
intermediate cusps to the inner \cite{NFW}\textbf{ }profile. Our
figures are illustrative however and the constants of our formula
would have to be chosen for the best fit consistent with positivity
in any particular case.

Figure (\ref{fig:regimes}) indicates these various regimes when the
constants are chosen such that the bulge mass is about fifteen (actually
$5\pi$, we represent the bulge mass by $M_{o}=4\pi\rho_{o}r_{o}^{3}$)
times that of the central black hole. Starting from upper left the
curves move inwards in a clockwise direction to show the outer $r^{-7/4}$
oscillating region, then the $r^{-1}$ region, a transition region,
and finally the $r^{-2}$ region. The curves are connected in a continuous
fashion.

\begin{figure}[t]
~\vspace{-1.5cm}

~\hspace{-0.501cm}\includegraphics[bb=20bp 218bp 575bp 633bp,width=1\columnwidth]{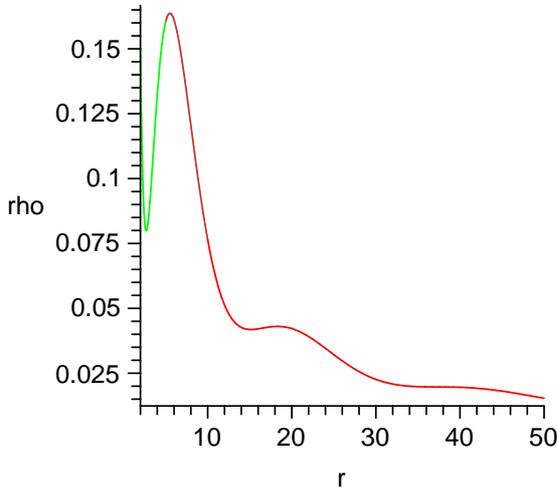}\vspace{1.6cm}

\caption[]{\label{fig:transition} This figure shows the oscillations in the
outer zero flux regime.}

\end{figure}

The final figure (\ref{fig:transition}) expands the outer oscillating
Bahcall-Wolf region.

Because of the slow decline in the density with radius, it is evident
that the outer radial limit to this solution is finite. Exactly where
it applies will depend on special numerical cases, but it seems in
any case that it must be inside the radius where the density has an
inverse square profile, essentially the scale radius of the dark matter
simulations.

The physical behaviour all stems from a DF that has the simple form
$K/|j|$. It appears to describe a zero flux equilibrium condition
in the mean, together with a stably filled loss cone (\cite{T2005}.
The density oscillations may indicate that collective behaviour is
necessary. It is an isolated example of the class of distribution
functions studied by Evans and An (\cite{EA2006}).

We can estimate the growth of a black hole in such a bulge in a straightforward
way according to \begin{equation}
\frac{dM_{\bullet}}{dt}=4\pi r_{\bullet}^{2}\rho_{\bullet}(\overline{v_{r}})_{\bullet},\end{equation}
 where $r_{\bullet}=6GM_{\bullet}/c^{2}$. The required mean radial
velocity is found to be $\overline{v_{r}}=4\sqrt{2}K/(3\rho r)|\Phi|^{3/2}$
which is the same for the inward or outward going particles in a steady
state. We calculate $\overline{v}_{r}$ from the expression \begin{align}
\overline{v}_{r} & =\frac{K}{r^{2}\rho}\int v_{r}dv_{r}\int\frac{dj^{2}}{(j^{2})^{1/2}},\end{align}
which becomes explicitly on transforming to an energy variable through
$dE=v_{r}dv_{r}$ \begin{align}
\overline{v}_{r} & =\frac{K}{r^{2}\rho}\int_{\Phi}^{0}dE\int_{0}^{2r^{2}(E\lyxmathsym{\textminus}\Phi)}\frac{dj^{2}}{(j^{2})^{1/2}}.\end{align}
This is conveniently written as \begin{align}
\overline{v}_{r} & =\frac{2K}{r^{2}\rho}\int_{0}^{\left|\Phi\right|}d\left|E\right|\sqrt{2r^{2}(\left|\Phi\right|\text{\textminus}\left|E\right|)},\end{align}
which yields finally the quoted result. In effect we have by ignoring
$dE=-v_{r}dv_{r}$ calculated only the outward mean velocity which
we take to be equal to the inward mean velocity of particles reaching
the black hole. We eliminate $K$ in terms of the bulge quantities
using equation (\ref{eq:nonSSdens}) so that \textbf{$2\pi K=r_{b}\rho_{b}/|\Phi|_{b}$}.
By setting $|\Phi|_{\bullet}=c^{2}/6$, the calculation yields finally
(restoring units)\begin{equation}
\frac{d\ln{M_{\bullet}}}{dt}=\frac{8}{\sqrt{27}}\left(\frac{r_{b}G\rho_{b}c}{|\Phi|_{b}}\right).\label{eq:BHgrowth}\end{equation}
 As an illustration we take $\rho_{b}$ to be the mean density of
$10^{9}M_{\odot}$ in one kiloparsec (i.e. $r_{b}$) and set $|\Phi|_{b}=v_{circ}^{2}(r_{b})$,
the circular velocity at $r_{b}$. We take this to be $200$ km/sec
and obtain finally \begin{equation}
\frac{d\ln{M_{\bullet}}}{dt}=1.3\times10^{-6}(yr)^{-1}.\label{eq:eFoldingBHtime}\end{equation}
 Given the e-folding time on the right of this last equation, we can
increase the mass of the black hole one hundred fold in three to four
million years. On this basis one can expect the presence of massive
black holes very early in the history of the Universe. This has to
happen however before the bulge has become isotropic, presumably in
the rapid growth phase of the dark matter halo. Some seed mass is
nevertheless required, which may be primordial.

Such a multi-power law behaviour does not give the simple linear correlation
between bulge mass and black hole mass that we suggested in papers
I and II. However since the black hole is incorporated into the global
solution at all times, we do expect such a correlation from the integration
of equation (\ref{eq:nonSSdens}) integrated over $r$. If most of
the inner part of the outer $r^{-7/4}$ law can be considered to have
fallen into the black hole, then $M_{\bullet}/M_{b}\approx(r_{\bullet}/r_{b})^{5/4}$. 

{}

\section{Conclusions}

We have sought in this series of papers (\cite{HLeDMcM09a,HLeDMcM09b}\textbf{
and the present paper}) to find distribution functions that describe
both dark matter bulges and a central black hole or at least a central
mass concentration. In most cases we succeed only in describing the
dark matter bulges, but there are some notable exceptions.

In the discussion of cusps and bulges based on purely radial orbits
(paper \cite{HLeDMcM09a}), we were able to distinguish the Distribution
function of Fridmann and Polyachenko (\ref{eq:FPDF}) from that of
Henriksen and Widrow (\ref{eq:steadyF}). The FPDF was found to describe
accurately the purely radial simulations of isolated collisionless
halos carried out in (\cite{MacM2006}). These simulations retained
the initial cosmological conditions although non-radial forces were
switched off. The final state is close to self-similar virialisation
rather than steady virialisation, since the infall continues. %
\begin{comment}
Moreover we showed in the last sub-section of the radial analysis
that the FPDF appears in the self-similar coarse graining at zeroth
order, if sensitivity to initial conditions is to be lost dynamically.
\end{comment}
{}

This correspondence between the theory and the simulations gives us
some confidence in the DF's found by remaining `close' to self-similarity.
This is especially so since predictions describing the simulation
results\textbf{ }in paper\textbf{ }\cite{HLeDMcM09b}\textbf{ }were
based on $a=0.72$ , which was deduced elsewhere in the context of
adiabatic self-similarity (\cite{H2007}).

The FPDF can contain consistently\textbf{ }a central mass concentration
that is unlikely to be a true black hole, at least in the early stages.
It may represent a central mass concentration or bulge initially.
%
\begin{comment}
The growth time of such a central mass in a system of radial orbits
is given simply by the dynamical time, so that this would be a rapid
phase.
\end{comment}
{} Subsequently with the rise of dissipation and instabilities, there
may be a slower phase of radial accretion towards the centre. It is
possible that this cycle could repeat several times in a process we
have referred to as `interrupted accretion'. Under this process the
$r^{-2}$ density law would apply almost everywhere. The radial velocity
dispersion is proportional to the potential. Thus it decreases as
$r^{-1}$ near the central mass, and subsequently decreases logarithmically
with $r$.

The HWDF (\ref{eq:steadyF}) is restricted to a strictly steady and
self-similar bulge, but it has the merit of allowing a family of densities
and potentials (velocity dispersion) according to the self-similar
prescription. A central mass is allowed only in the Keplerian limit
wherein $a=3/2$. This gives a massless bulge with $\rho\propto r^{-3}$.
This is naturally iterated to give an inner flattening but continued
iteration is effectively in powers of $\ln{r}$, which should therefore
yield weak corrections. The iteration can only apply away from $r=0$,
so that the central mass is a `renormalized' extended mass.

The HWDF was shown in paper \cite{HLeDMcM09a} to give a density that
is linear in the potential\textbf{,} and hence a self-consistent bulge
is found from the Poisson equation. The density profile is never flatter
than $r^{-2.5}$ near the central mass and tends to $r^{-3}$ in the
near Keplerian limit of dominant central mass. This restricts the
applicability to a region outside the central bulge. It does not seem
to be relevant to a near black hole domain.

%
\begin{comment}
Our final result concerning steady, self-similar radial orbits concerned
the special case $a=1$. The DF is a Gaussian that has been found
previously in coarse graining. We include it here as a second example
of a radial DF that produces an $r^{-2}$ density profile (\cite{Mutka09}),
although with logarithmic corrections. It is not strictly self-similar.
\end{comment}
{}

The inclusion of angular momentum led to more realistic situations.
We re-derived in paper\textbf{ }\cite{HLeDMcM09b} the steady self-similar
DFs from first principles in equation (\ref{eq:anisopsteady}). We
showed that these can be used to describe the simulated collisionless
halos calculated in (\cite{MacM2006}), if we use the value of $a\approx0.72$
identified in (\cite{H2007}) and take two limits. In one (\ref{eq:DFE})
the DF is isotropic and describes approximately the central region
of the bulge. The other limit (\ref{eq:DFJ}) describes the outer
region. This encourages us regarding the relevance of this family.
Unfortunately the potential of a central mass can not be included
exactly in these distribution functions. %
\begin{comment}
Iteration is possible along the lines discussed for radial bulges,
but since the growth of a black hole in an isotropic central bulge
is negligible, we have not pursued this avenue.

We have also found a self-similar generalization of the radial FPDF
in equation (\ref{eq:genFPDF}). Unfortunately this DF does not have
the property of yielding a density that is independent of the potential
and so a point mass potential is incompatible with self-similarity.
It might describe a system at large radii although the description
there is unlikely to be unique.
\end{comment}
{}

As always $a=1$ must be treated separately and we give a derivation
from first principles in this series. In paper \cite{HLeDMcM09a}
we show that this case corresponds to a radially growing system with
the FPDF. The anisotropic solution is discussed in paper \cite{HLeDMcM09b}
The results are new. It implies the density profile%
\begin{comment}
. The result is new. It does generalize the FPDF in the sense that
\end{comment}
{} $\rho\propto r^{-2}$ always. %
\begin{comment}
Although we can not place a central mass inside this bulge exactly,
it allows an anisotropic DF in the $r^{-2}$ bulge region just as
does the DF of the preceding paragraph..
\end{comment}
{}

Generally we find that we can not describe elegantly anisotropic bulges
containing black holes with self-similar DFs, as one might expect.
The self-similarity is restricted to the surrounding bulge, and this
is progressively perturbed near the black hole. We re-emphasized in
the current paper in a non self-similar context that a sharp cut-off
of any kind (whether due to black hole binary scouring or not) can
yield a density profile as flat as $-1/2$.

Our most successful description of a black hole embedded in an anisotropic
bulge is given in this paper by the DF $\pi f=K/(j^{2})^{1/2}$. This
simplest member of the family discovered by Evans and An (ibid) yields
a bulge containing a central point mass. The density profile near
this mass is given by equation (\ref{eq:CuspR1}) as $r^{-1}$ with
an inner $r^{-2}$ peak. Farther out there is an $r^{-7/4}$ domain
in the mean that imitates the Bahcall and Wolf (ibid) cusp, which
in turn ultimately becomes the NFW $r^{-1}$ profile.

The reasons for this behaviour appear to be very different from those
of Bahcall and Wolf, since there are no two body collisions in this
treatment. By assuming a steady filled loss cone of this simple form,
we have selected a zero flux behaviour on the outer boundary as an
average behaviour%
\begin{comment}
everywhere except at the black hole boundary
\end{comment}
{} independently of the detailed mechanism that acts to establish this
($\partial f/\partial E=0$). In that sense it is a useful description
of the black hole cusp region. The oscillations may indicate the necessity
of collective behaviour for maintaining a filled loss cone. 

We observe finally that this DF does not require a very strong concentration
of low angular momentum particles. In the outer region where $\Phi\approx constant$,
we find $dN/dj^{2}\approx constant$ so that the particles are mostly
at high angular momentum, as one might expect.

Black hole growth is astrophysically rapid in this distribution as
seems to be required observationally. Careful study of the density
regimes may permit bulge mass and black hole mass to be distinguished
and correlated, assuming this distribution is realized.

\section{Acknowledgements}

RNH acknowledges the support of an operating grant from the canadian
Natural Sciences and Research Council. The work of MLeD is supported
by CSIC (Spain) under the contract JAEDoc072, with partial support
from CICYT project FPA2006-05807, at the IFT, Universidad Autonoma
de Madrid, Spain%{Zel'dovich \& Podurets, 1965}
%\bibitem[Bertschinger, 1985]{Bertschinger85}Bertschinger, E., 1985, ApJS, 58, %39. 
%\bibitem[Del Popolo {\it et al}, 2000]{DelPopolo00}Del Popolo, A., Gambera, M.%, Rercami, Spedicato, E., 2000, A\&A, 353,
%427.

\end{comment}
{}

\bibitem[HW95]{HW95}Henriksen, R.N., Widrow, L.M., 1995, MNRAS, 276,
679.

\bibitem[HW1997]{HW97}Henriksen, R.N., Widrow, L.M., 1997, Phys.
Rev. Lett., 78, 3426.

\bibitem[HW 1999]{HW99}Henriksen, R.N., Widrow, L.M., 1999, MNRAS,
302, 321.%
\begin{comment}

\end{comment}
{}

\bibitem[H2006A]{H2006A} Henriksen, R.N.,2006, MNRAS, 366, 697.

\bibitem[H2006]{H2006}Henriksen, R.N., 2006, ApJ, 653,894.

\bibitem[H2007]{H2007} Henriksen, R.N., 2007, ApJ,671,1147.%
\begin{comment}
\bibitem[H2009]{H2009} Henriksen, R.N., 2009, ApJ, 690, 102.
\end{comment}
{}

\bibitem[Kurk et al., 2007]{Kurk2007} Kurk, J.D., et al., 2007, ApJ,
669, 32.

%\bibitem[Jing \& Suto, 2000]{JingSuto}Jing, Y.P., Suto, Y., 2000, ApJ, 529, L6%9.

\bibitem[KR1995]{KR1995}Kormendy, J.,\& Richstone, D., 1995, Ann.Rev.A\&A.%
\begin{comment}
\bibitem[Kormendy \& Bender 2009]{KB2009}Kormendy, J., \& Bender,
R., 2009, ApJ, 691,L142.
\end{comment}
{}

%\bibitem[Kravtsov {\it et al} 1998]{Kravtsov98}Kravtsov, A.V., Klypin, A.A., %Bullock, J.S., Primack, J.R., 1998,
%ApJ, 502, 48.

\textbf{\bibitem[I]{HLeDMcM09a}}Le Delliou, M., Henriksen, R.N.,
\& MacMillan, J.D., 2010, subm. to MNRAS {[}arXiv : 0911.2232{]} (I)

\bibitem[II]{HLeDMcM09b}Le Delliou, M., Henriksen, R.N., \& MacMillan,
J.D., 2010, accepted by A\&A {[}arXiv : 0911.2234{]} (II)\textbf{}%
\begin{comment}
\textbf{\bibitem[III]{HLeDMcM09c}Le Delliou, M., Henriksen, R.N.,
\& MacMillan, J.D., 2009, subm. to A\&A {[}arXiv : 0911.2238{]} (III)}
\end{comment}
{}

\bibitem[Le Delliou 2001]{LeD2001}Le Delliou, M., 2001, PhD Thesis,
Queen's University, Kingston, Canada.%
\begin{comment}

\end{comment}
{}

\bibitem[MacMillan 2006]{MacM2006}MacMillan, J., 2006, PhD Thesis,
Queen's University at Kingston, ONK7L 3N6, Canada.

\bibitem[MWH 2006]{MWH2006} MacMillan, J.D., Widrow, L.M., \& Henriksen,
R.N., 2006, ApJ,653, 43.

\bibitem[Ma98]{Ma98}Magorrian, J., et al., 1998, AJ, 115, 2285.

\bibitem[Maiolino et al. 2007]{Mai2007}Maiolino, R., et al.,2007,
A\&A, 472, L33.

\bibitem[Merritt \& Szell 2006]{MS2006} Merritt, D., \& Szell, A.,
2006, ApJ, 648, 890.

\bibitem[Mutka 2009]{Mutka09} Mutka, P., 2009, Proceeding of \textit{Invisible
Universe}, Palais de l'UNESCO, Paris, ed. J-M Alimi.

\bibitem[Peirani \& de Freitas Pacheo 2008]{PFP2008}Peirani,S. \&
de Freitas Pacheo, J.A., 2008, Phys. Rev. D, 77 (6), 064023.

%\bibitem[Merrit {\it et al}, 1989]{MTJ}Merritt, D., Tremaine, S., Johnstone, D%., 1989, ApJ, 236, 829.

\bibitem[Nakano \& Makino 1999]{NM99}Nakano, T., Makino, M., 1999,
ApJ, 525, L77.

\bibitem[NFW]{NFW}Navarro, J.F., Frenk, C.S., White, S.D.M., 1996,
ApJ, 462, 5%63.
%
\begin{comment}

\end{comment}
{}

%\bibitem[Ryden, 1988]{Ryden88}Ryden, B.S., 1988, ApJ, 333, 78.
%\bibitem[Shapiro {\it et al}, 1986]{Shapiro86}Shapiro, S., Teukolsky, S.A., 19%86, ApJ, 307, 575.
%\bibitem[Sikivie {\it et al}, 1997]{Sikivie97}Sikivie, P., Tkachev, I.I., Wang%, Y., 1997, Phys. Rev. D, 56.
%\bibitem[Stiavelli \& Bertin, 1985]{StiavelliBertin}Stiavelli, M., Bertin, G., %1985, MNRAS, 217, 735.
%\bibitem[Stil, 1999]{Stil99}Stil, J., 1999, Doctoral Thesis, Leiden Observatory.
%\bibitem[Subramanian 2000]{Subra00}Subramanian, K., 2000, ApJ, in press

\bibitem[Tremaine 2005]{T2005}Tremaine, S., 2005, ApJ, 625,143.%
\begin{comment}

\end{comment}
{}
\end{thebibliography}


\begin{thebibliography}{Tremaine 2005}
\bibitem[Baes et al. (2005)]{Baes05}Baes, M., Dejonge, H., \& Buyle,P.,
A\&A, 432, 411 (2005).

\bibitem[Bahcall \& Wolf 1976]{BW76} Bahcall,J.,\& Wolf, R.A., 1976.
ApJ, 209, 214. 

\bibitem[Binney \& Tremaine 1987 ]{BT1987} Binney, J. \& Tremaine,
S., 1987. \textit{Galactic Dynamics}, Princeton University Press,
Princeton, New Jersey.

\bibitem[Carter \& Henriksen 1991]{CH91}Carter,B. \& Henriksen, R.N.,
1991, J. Math. Phys., 32, 2580.

\bibitem[Evans \& An 2006]{EA2006}Evans,N.Wyn.,\& An, Jin H., 2006,
Phys. Rev. D, 73(2), 023524.%
\begin{comment}
\bibitem[Diemand et al.06]{DKM2006}Diemand, J., Kuhlen,M. \& Madau,
P., 2006,ApJ, 667,859.
\end{comment}
{}

\bibitem[Ferrase and Merritt 2000]{FM2000}Ferrarese,L.,\& Merritt,
D., 2000, ApJ,539, L9.


%\bibitem[Fillmore \& Goldreich, 1984]{FG84}Fillmore, J. A., Goldreich, P., 198%4, ApJ, 281, 1.


\bibitem[Fridmann \& Polyachenko 1984]{FP1984}Fridman,A.M., \& Polyachenko,
V.L., 1984,\textit{Physics of Gravitating Systems}, Springer, New
York.

\bibitem[Fujiwara 1983]{Fujiwara}Fujiwara, T., 1983, PASJ, 35, 547.

\bibitem[Gebhardt et al 2000]{Geb2000}Gebhardt,K., et al.,2000, ApJ,539,L13.%
\begin{comment}
\begin{thebibliography}{Tremaine 2005}
\bibitem[Gillessen et al. 2009]{G2009} Gillessen, S., et al., 2009,
ApJ, 692, 1075.

\bibitem[Hansen 2004]{Hansen2004} Hansen, S., 2004, MNRAS, 352, L41.
\end{thebibliography}

\begin{thebibliography}{Tremaine 2005}
\bibitem[HLeD 2002]{HLeD2002}Henriksen, R.N., \& Le Delliou, M.,
2002, MNRAS, 331, 423.

\bibitem[H2004]{H2004} Henriksen, R.N., 2004, MNRAS, 355, 1217.
\end{thebibliography}

\begin{thebibliography}{Tremaine 2005}
\bibitem[Lu et al. 2006]{Lu2006} Lu, Yu, 2006, MNRAS,368, 193.

\bibitem[MacMillan \& Henriksen 2002]{MH2002} MacMillan, J.D., \&
Henriksen, R.N., 2002, ApJ, 569,83.
\end{thebibliography}

\begin{thebibliography}{Tremaine 2005}
\bibitem[Peebles 1972]{P1972}Peebles, P.J.E., 1972, Gen.Rel.Grav.,
3, 63.

\bibitem[Quinlan et al 1995]{Q1995}Quinlan, G.D., Hernquist, L.,
Sigurdsson, S., 1995, ApJ, 440, 554.
\end{thebibliography}

\begin{thebibliography}{Tremaine 2005}
\bibitem[van der Marel 2009]{vanderMarel09} van der Marel, R., 2009,
in \textit{Unveiling the Mass}, Queen's U., Kingston, Ontario, ed.
S. Courteau.

\bibitem[Young 1980]{Y1980}Young, P., 1980, ApJ, 242, 1232.

\bibitem[Zhao et al., 2003]{Zhao2003} Zhao, D.H. et al., 2003, MNRAS,
339. 12.


%\bibitem[Zel'dovich \& Podurets, 1965]{Zel'dovich65}Zel'dovich, Ya.B;, Poduret%s, M.A., 1965, Soviet Astr.-A.J., 9, 742.


\bibitem[Walter et al., 2009]{W09} Walter, F., et al., 2009, Nature,457,
699.
\end{thebibliography}
\end{document}